%% file: main.tex
\documentclass[]{article}

\makeatletter\if@twocolumn\PassOptionsToPackage{switch}{lineno}\else\fi\makeatother

\usepackage{amsfonts,amssymb,amsbsy,latexsym,amsmath,tabulary,graphicx,times,caption,fancyhdr}
\usepackage[utf8]{inputenc}
\usepackage{url,multirow,morefloats,floatflt,cancel,tfrupee}
\makeatletter

\AtBeginDocument{\@ifpackageloaded{textcomp}{}{\usepackage{textcomp}}}
\makeatother
\usepackage{colortbl}
\usepackage{xcolor,soul}
\usepackage{pifont}
\usepackage[nointegrals]{wasysym}
\makeatletter

\def\mcWidth#1{\csname TY@F#1\endcsname+\tabcolsep}

\def\cAlignHack{\rightskip\@flushglue\leftskip\@flushglue\parindent\z@\parfillskip\z@skip}
\def\rAlignHack{\rightskip\z@skip\leftskip\@flushglue \parindent\z@\parfillskip\z@skip}

\@ifundefined{etal}{}{}

\usepackage{ifxetex}
\ifxetex\else\if@twocolumn\@ifpackageloaded{stfloats}{}{\usepackage{dblfloatfix}}\fi\fi

\AtBeginDocument{
\expandafter\ifx\csname eqalign\endcsname\relax
\def\eqalign#1{\null\vcenter{\def\\{\cr}\openup\jot\m@th
  \ialign{\strut$\displaystyle{##}$\hfil&$\displaystyle{{}##}$\hfil
      \crcr#1\crcr}}\,}
\fi
}

\AtBeginDocument{%
  \@ifpackageloaded{endfloat}%
   {\renewcommand\efloat@iwrite[1]{\immediate\expandafter\protected@write\csname efloat@post#1\endcsname{}}}{\newif\ifefloat@tables}%
}%

\def\BreakURLText#1{\@tfor\brk@tempa:=#1\do{\brk@tempa\hskip0pt}}
\let\lt=<
\let\gt=>
\def\processVert{\ifmmode|\else\textbar\fi}

\@ifundefined{subparagraph}{
\def\subparagraph{\@startsection{paragraph}{5}{2\parindent}{0ex plus 0.1ex minus 0.1ex}%
{0ex}{\normalfont\small\itshape}}%
}{}

\newcommand\role[1]{\unskip}
\newcommand\aucollab[1]{\unskip}
  
\@ifundefined{tsGraphicsScaleX}{\gdef\tsGraphicsScaleX{1}}{}
\@ifundefined{tsGraphicsScaleY}{\gdef\tsGraphicsScaleY{.9}}{}
\def\checkGraphicsWidth{\ifdim\Gin@nat@width>\linewidth
	\tsGraphicsScaleX\linewidth\else\Gin@nat@width\fi}

\def\checkGraphicsHeight{\ifdim\Gin@nat@height>.9\textheight
	\tsGraphicsScaleY\textheight\else\Gin@nat@height\fi}

\def\fixFloatSize#1{}
\let\ts@includegraphics\includegraphics

\def\inlinegraphic[#1]#2{{\edef\@tempa{#1}\edef\baseline@shift{\ifx\@tempa\@empty0\else#1\fi}\edef\tempZ{\the\numexpr(\numexpr(\baseline@shift*\f@size/100))}\protect\raisebox{\tempZ pt}{\ts@includegraphics{#2}}}}

\AtBeginDocument{\def\includegraphics{\@ifnextchar[{\ts@includegraphics}{\ts@includegraphics[width=\checkGraphicsWidth,height=\checkGraphicsHeight,keepaspectratio]}}}

\DeclareMathAlphabet{\mathpzc}{OT1}{pzc}{m}{it}

\def\URL#1#2{\@ifundefined{href}{#2}{\href{#1}{#2}}}

\def\UrlOrds{\do\*\do\-\do\~\do\'\do\"\do\-}%
\g@addto@macro{\UrlBreaks}{\UrlOrds}

\edef\fntEncoding{\f@encoding}

\makeatother

\newif\ifmultipleabstract\multipleabstractfalse%
\makeatletter

\def\wileyIndent{1pt}
\usepackage[paperheight=10in,paperwidth=6.5in,margin=2cm,headsep=.5cm,top=2.5cm]{geometry}

\renewenvironment{abstract}
{\vspace*{-1pc}\trivlist\item[]\leftskip\wileyIndent\hrulefill\par\vskip4pt\noindent\textbf{\abstractname}\mbox{\null}\\}{\par\noindent\hrulefill\endtrivlist}

\usepackage[]{footmisc}

\def\author#1{\gdef\@author{\hskip-\dimexpr(\tabcolsep)\hskip\wileyIndent\parbox{\dimexpr\textwidth-\wileyIndent}{\centering\bfseries#1}}}

\def\title#1{\linespread{1}\gdef\@title{\centering\bfseries\ifx\@articleType\@empty\else\@articleType\\\fi#1}}

\let\@articleType\@empty \def\articletype#1{\gdef\@articleType{{\normalfont\itshape#1}}}

\AtBeginDocument{\fancypagestyle{headings}{\fancyhf{}\fancyhead[C]{\RunningHead}\fancyhead[R]{\thepage}}\pagestyle{headings}}

 \def\audegree#1{}

\date{}

\makeatother

\usepackage[T1]{fontenc}
\makeatother
\usepackage[authoryear]{natbib}

\def\thanksspace{{\phantom{\textsuperscript{\thefootnote}}}}

\usepackage{float}
\usepackage{siunitx}
\usepackage{longtable}
\usepackage{hhline}
\usepackage{booktabs}

\begin{document}

\title{Mind the wealth gap: a new allocation method to match micro and macro statistics for household wealth}
\author{Michele~Cantarella,\textsuperscript{1}\thanks{Corresponding author. E-mail:
                    michele.cantarella@helsinki.fi}{\thanksspace}
                    Andrea~Neri\textsuperscript{2} and Maria Giovanna~Ranalli\textsuperscript{3}~\\[-3pt]\normalsize\normalfont  \itshape ~\\
\textsuperscript{1}{Centre for Consumer Society Research\unskip, University of Helsinki}~\\
\textsuperscript{2}{Economic and Financial Statistics Department \unskip, Banca d'Italia}~\\
\textsuperscript{3}{Department of Political Science\unskip, Universit\`a degli Studi di Perugia\unskip}}

\def\RunningHead{}\def\RunningAuthor{Cantarella, Neri and Ranalli}

\maketitle 

\begin{abstract}
The financial and economic crisis recently experienced by many European countries has increased demand for timely, coherent and consistent distributional information for the household sector. In the Euro area, most of the national central banks collect such information through income and wealth surveys, which are often used to inform their decisions. These surveys, however, may be affected by non-response and under-reporting behaviours which leads to a mismatch with macroeconomic figures coming from national accounts.  In this paper, we develop a novel allocation method which combines information from a power law (Pareto) model and imputation procedures so to address these issues simultaneously, when only limited external information is available. Finally, we produce distributional indicators for four Euro-Area countries.

\def\keywordstitle{Keywords}


\end{abstract}

\section{Introduction}
\label{section:intro}

The financial and economic crisis recently experienced by many European countries has increased demand for timely, coherent, and consistent distributional information relating to household income and wealth. Such information is receiving high priority especially in the agenda of national central banks (NCBs) which use it in several ways \citep{hfcs2009}. Distributional information is used for financial stability purposes, for example, to assess how much debt is concentrated in the hands of financially vulnerable households \citep[see, for instance,][]{AMPUDIA2016, qef_369_16}. Moreover, distributional information allows to estimate the aggregate consumption response to wealth shocks when individual responses are heterogeneous \citep{Pa2007,GPV} and, more generally, to understand the interplay between monetary policy measures, especially non-standard ones, and the distribution of income and wealth \citep{cas,joes, COIBION201770}. Recent years have also been characterised by a surge of interest in the study of the dynamics of wealth accumulation over the last century \citep{gp2018,Alvaredo2018,Frmeaux2020}.

Sample surveys are the main source of distributional information on household wealth.  
In the Euro area, most NCBs conduct the Eurosystem Household Finance and Consumption Survey (HFCS) which collects harmonized household-level data on households’ finances and consumption \citep{hfcs2009}. 

The second source of information relating to household wealth comes from national accounts 
which record the stock of assets, both financial and non-financial, and liabilities at a particular point in time.  

In theory, since the HFCS is designed to be representative of all households, aggregating this microdata should correspond to the macro aggregates. In practice, however, differences are large: aggregate totals based on surveys are often substantially below the totals to be found in national accounts. Before using the distributional information from survey data, it is, therefore, crucial to explain and possibly eliminate the differences between the two sources of information.\footnote{In 2015, the European System of Central Banks (ESCB) has established an expert group with the aim of comparing and bridging macro data (i.e. national accounts/financial accounts) and microdata (i.e. the Household Finance and Consumption Survey) on wealth.}

There are several possible reasons for the differences \citep{egdna}. 
From the survey side, two relevant issues are unit non-response and measurement errors. There is substantial evidence that household's decision whether or not to participate in the survey is not at random. In particular, wealthy households are  difficulty to contact and convince to participate \citep{Chakraborty19,AK2008,Ken19,vermeulen2018fat}. Since these households own a large share of total wealth, their under-representation in the final sample is likely to result in a biased picture of the wealth distribution.

Moreover, wealth surveys generally include both complex and sensitive items. As a consequence, respondents are not always able or even willing to report the correct amount of wealth they hold. Similar to non-response, measurement error is not at random and differs across population subgroups and portfolio items \citep{Neri15,Ran11}.

The ideal solution for overcoming these problems would be to link survey data with administrative records \citep[such as tax records or credit registers, as in][]{blanchetetal2018,gp2018,garbinti2017accounting}. Alternative approaches to data linkage are directly based on the use of wealth (tax) records \citep{Alva2009} or on the use of capital income information from tax records to construct wealth estimates assuming certain rates of return on wealth \citep{saez16}.  

Unfortunately, when such administrative records exist and are not limited in scope, they are not usually available for confidentiality reasons. 
Because of that, the recent literature has developed methods to combine survey data with the limited external information publicly available, such as aggregate figures from national accounts or lists of rich individuals' total wealth. 
\citet{vermeulen2018fat,Vermeulen2016} uses Forbes World Billionaires lists in combination with some wealth surveys to estimate the total wealth held by rich households. He shows that the use of such lists increases the quality of the results (compared to estimating a Pareto model from survey data alone). 
Building on this approach, \citet{Chakraborty19}, \citet{Waltl2018}, and \citet{chakraborty2018missing} extend the analysis by benchmarking survey results to the national accounts. In another recent study, \citet{Bach2019} implement these methodologies to impute rich list data to wealth surveys.

The common assumption behind all these studies is that unit non-response of wealthy households is the only reason for the micro-macro gap.

This paper contributes to the literature which tries to produce distributional indicators of wealth that are consistent with the national accounts, by proposing a methodology that draws on existing and well-established methods.We contribute to this literature in four ways. 

First, whereas previous studies focus only on the missing part of the tail, assuming existing survey observations as representative, we claim that differential non-response also affects the representativeness of existing survey observations, which in turn affects estimates for the total number of households in the Pareto tail, and their total wealth. We propose a correction for differential non-response that accounts for the missing rich, but focuses on observed survey households.
By any means, this correction does not substitute the imputation \citep{Bach2019} or simulation \citep{Waltl2018} procedures developed in the literature, but rather complements them allowing for the correction of non-response bias among existing survey observations.

Second, while existing papers only focus on non-response at the tail of the distribution, we present a methodology that allows us to correct also for measurement error. Dealing with both aspects simultaneously is important, even when the research purpose is to estimate the share of total wealth held by wealthy households. Indeed, some rich households may misreport their true wealth and therefore they could be misclassified in adjustment process. An advantage of our approach is that it enables us to compute distributional indicators that refer to "non-rich" households, such as those relating to financial vulnerability.   

Our third contribution is that even if we apply methods that are well-established (such as the Pareto model, imputation, and calibration) we show how to combine and use them in a single framework and how to evaluate the precision of the results. 

The fourth contribution is to produce a modified and readily usable dataset in which survey values are adjusted for the above-mentioned quality issues and, by construction, the totals add up to the national accounts. While the existing papers are mainly focused on methods to estimate of total wealth held at the top, our adjusted dataset can be used for estimating any distributional indicator that may be of interest.

The paper is structured as follows. Section \ref{section:data} describes the data sources used in our application and motivating example. Section \ref{section:Methodology} presents the Pareto approach (section \ref{subsection:ParetoTail}) and calibration (section \ref{subsection:SvyCalibration}) and the methodologies we use to combine them in a single approach (sections  \ref{subsection:parcal}, \ref{subsection:simultaneous},
\ref{subsection:wealthcalib}).
Section \ref{section:assessment} describes the tools used to assess the properties of the proposed methods. 
Section \ref{section:application} describes how the method is applied to our data, while section \ref{section:results} discusses the results and the main findings of the application. Section \ref{section:conclusions} provides some conclusions and lines for future research.

\section{Data}
\label{section:data}
This paper uses the Household Finance and Consumption Survey (HFCS) and two sources of auxiliary information, that is the national accounts which include both financial and non-financial accounts, and rich list data. 

The Household Finance and Consumption Survey (HFCS) is a joint project of all the national central banks (NCBs) of the Eurosystem and several national statistical institutes (NSIs). The survey collects detailed household-level data on various aspects of household balance sheets and related economic and demographic variables, including income, private pensions, employment, and measures of consumption. The HFCS is conducted in a decentralised manner.   A group of experts from the European Central Bank (ECB) and from the NCBs (the Household Finance and Consumption Network, HFCN) coordinates the whole project, ensuring the cross-country comparability of the final data. 

We use the second wave of HFCS (2014) and we restrict our analysis to four countries: Italy, France, Germany, and Finland. This choice is motivated by two considerations. First, rich lists and non-financial accounts are available for this subset of countries. Second, these surveys present methodological differences that can be used to evaluate our method. For example, some countries over-sample rich households using individual tax records (as in the French and Finnish survey) or using the information at the regional level (as in the German one), while others do not over-sample (as in the Italian case). Moreover, in some cases, the survey is linked with administrative data (as in the Finnish one). In both cases of over-sampling and use of administrative records, we should expect a lower effect of the adjustment method.   

Our variable of interest is household net wealth defined as the sum of deposits, bonds, shares, mutual funds, money owed to the household, the value of insurance policies and pension funds, business wealth, and housing wealth, minus debts. 

The second source of information is national accounts. The financial component (financial accounts) is produced by NCBs and relates the total financial assets and liabilities held by households, classified by financial instrument, in order of liquidity based on the original maturity and negotiability (from cash to deposits and insurance and pension instruments).
Non-financial accounts are produced by NSIs and contain the total value of dwellings, other buildings and structures, and land owned by households. 
Even if national accounts figures may suffer from quality issues and may adopt different concepts and definitions from the ones used in the survey, we use them as a benchmark to correct survey data.   

Rich lists are our third source of information. They have already been used in the literature to adjust for missing rich households \citep{vermeulen2018fat, chakraborty2018missing}.  Their use may generate concerns since the methodology adopted is often obscure and usually only figures for net worth are provided, with no financial instrument breakdown. Some studies have tried to overcome these issue by using different types of Pareto adjustments  \citep{blanchet2017generalized, Waltl2018}. 
Other studies \citep[such as][]{Schrder2019} have also explored new ways of sampling high-wealth individuals with adequate precision. However, these methods can only be employed in specific instances when information on these households exists and is easily accessible. When these sources are not available, rich lists remain a reliable alternative, and evidence from \citet{Waltl2018} indicates that, after the integration with rich lists, there might be little difference between the wealth estimated by different Pareto adjustments.

In our case, we use wealthy household data from the 2014 \textit{Forbes' Billionaires List}. This information has been replaced by that from larger region-specific lists, such as 2014 editions of \textit{Challenges' "Les 500 plus grandes fortunes de France"} for France, \textit{Manager Magazin}'s list for Germany and \textit{Arvopaperi}'s list for Finland, when available. We also adjust this rich list data by estimating the debts and portfolio composition, based on portfolio shares from top wealth observations in the HFCS.\footnote{This is a simplifying assumption. An improvement over this form of portfolio allocation can be offered by the approach used in \citet{chakraborty2018missing}.}  

In this way, estimates for portfolio compositions among top fortunes can be obtained, and rich list data can be fully integrated with the HFCS for estimation purposes.

\section{Methodology}
\label{section:Methodology}
 
Let $w$ be household net wealth and $t(w)$ the population total to be estimated using survey data. 
Let $\hat{t}(w)=\sum_{i=1}^{S}d_{i} w_{i}$  be the Horvitz-Thompson estimator, where $d_{i}$ is the sampling weight and $w_{i}$ the net wealth for each individual household $i$ in the sample of respondents $\mathcal{S}_0 = \{1,2,..., S\}$, ordered by net wealth rank.

Because of unit non-response and measurement error the expected value of the Horvitz-Thompson estimator $\hat{t}(w)$ is generally lower than  $t(w)$, the corresponding macro figure. 
Unit non-response occurs when some households refuse to participate to the survey. If this decision is related to household wealth (i.e richer households are more difficult to enrol in the survey than others) the sample of respondents $\mathcal{S}_0$ may not represent adequately the upper tail of the distribution. Measurement error happens when the information collected in the survey $w$ is different from the true unknown value $w^*$. The error term $(w^*-w)$ may depend on many factors such as the difficulty of respondents to recall the required information or their unwillingness to report their true wealth. 

Our methodology to address these issues is based on two techniques that are well-established in the literature.
We use the Pareto distribution to compensate for unit non-response of wealthy households (section \ref{subsection:ParetoTail}), and the calibration methods commonly used in survey sampling to deal with the issue of measurement error (section \ref{subsection:SvyCalibration}). 

The two correction methods are dependent on each other and they must be implemented simultaneously. The Pareto correction starts with an assessment of the rich households available in the survey. Because of measurement error, some households could be misclassified and therefore a preliminary calibration adjustment is required. 
On the other hand, calibration is used again for the adjustment for measurement error across the whole distribution, requiring that the survey represents adequately the upper tail of the distribution. 

Our solution to conduct the two adjustment simultaneously is to run them in an iterative process, based on the procedure described in the following sections. 

The final product of the methodology is an adjusted survey data set with total estimates of net wealth, real assets, financial assets and liabilities that match the aggregate figures in the national accounts balance sheet. This data set can be used to compute several distributional indicators of interest.

Before applying the method, we reclassify some definitions of wealth items used in the survey data in order to remove as many of the conceptual differences with national accounts as possible \citep[see for instance][]{eglmm,Chakraborty19}. In particular, we remove from national accounts totals the wealth held by non-profit institutions serving households (NPISHs), and we only focus on  the items with the highest level of comparability.

\subsection{Pareto tail estimation}
\label{subsection:ParetoTail}

The Pareto adjustment assumes that, over a certain wealth threshold $(w_{0})$, the complementary cumulative distribution (CCDF) of wealth is approximated by a power law, which (for $w_{i} \geq w_{0}$) can be expressed as:
\begin{equation}
\label{EQ:paretoCDF}
P(W\leq w_i)=1-(w_{0}/w_{i})^\alpha 
\end{equation}
where the parameter $\alpha \in \mathbb{R}^+$ indicates the shape of the tail. The lower the value of $\alpha$, the fatter is the tail,
and the more concentrated is wealth.

The first step of the adjustment is the estimation of the threshold $w_{0}$. Previous research has often adopted the arbitrary threshold of €1 million and, as a robustness check, of 1.5 or 2 million.
We relax this assumption by using a less arbitrary  method, based on the properties of the mean excess function \eqref{EQ:meanexcess} \citep{Yang1978}:
\begin{equation}
\label{EQ:meanexcess}
    E[W - w_{i} | W > w_{i}] = \frac{\sum_{j = 1}^{i} d_{j} (w_{j} - w_{i})}{\sum_{j = 1}^{i}  d_{j}}
\end{equation}
with $j \leq i$.
The expectation expressed by this function is estimated with the weighted mean of the deviation from $w_i$ for all observations $j$ whose wealth exceeds $w_i$. Essentially, every value of $w_i$ is treated as a possible threshold when the corresponding expected value of $E[W - w_{i} | W > w_{i}] $ is estimated.

A useful property of this function is its linearity in $w_{i}$ if the distribution is Pareto \citep{Yang1978,DavisonSmith1990}. Following from this property, we estimate $E[W - w_{i} | W > w_{i}]$ for each value of $w_{i}$ in $\mathcal{S}_0$, and then we find the threshold after which the mean excess function is linear on $w_i$. This can be achieved by selecting the value $w_{0}^*$ for which the R-squared of the linear regression of $E[W - w_{i} | W > w_{i}]$ on $w_{i}$ is maximised \citep{Langousis2016}.  

It is worth stressing that the threshold $w_0$ is the point where the Pareto distribution starts, which differs from the truncation point ($w_1$) after which the survey has no rich households. Indeed, survey data will generally include observations in the bottom part of the Pareto while missing those at the very top of the distribution. In the presence of truncation, the relationship between mean excesses and wealth will turn to take a downward bias the closer we approach the truncation point \citep[see][]{Aban2006}. To account for this issue, we weight the regression $E[W - w_{i} | W > w_{i}]$ on $w_{i}$ by the sum of survey weights for all $j \leq i$.\footnote{As a robustness check, we also used the 1 million threshold to estimate the Pareto shape parameter and ran all our adjustment methods afterward. These conservative estimates are very close to the ones obtained using our estimated threshold and are available on request.}

After the threshold $w_0$ is found, the shape parameter $\alpha$ can be estimated using the method described in \citet{vermeulen2018fat}.

Define $\mathcal{S}_T=\{1_T,2_T,...,m_T\}$ as the sub-sample of respondents with wealth higher than $w_0$.   
The rich list $\mathcal{S}_R=\{1_R,2_R,...,m_R\}$ and the sample $S_T$ are appended creating a new file $\mathcal{S}_I=\{1_I,2_I,...,m_I\}$ with $m_I$ observations. For simplicity, we will drop the sample subscript from now on. Households are again ordered by wealth rank $i$ where the lower the rank the higher household wealth. So the rank of the richest household in the sample is one, the rank for the second richest is two, and so on until household $m$, whose wealth $w_m$ equals the threshold $w_0$, is reached. 

Survey weights are taken into account by assigning to observations in the rich list weight $d_i = 1$, while survey observations (a subset of the sample $\mathcal{S}_0$) retain their original survey weight.  
Denote by $\bar{D}$ the average survey
weight of all observations in sample $S_I$ (i.e. $\bar{D} = \sum_{j=1}^{m}d_j/m$). Denote the sum of all weights as $D = \sum_{j=1}^{m}d_j$, representing an estimate of the number of households that have wealth at least as high as $w_0$. 
Define 
$\bar{D}_i$ the average weight of the first $i$ sample points (i.e. $\bar{D}_i =\sum_{j=1}^{i}d_j/i$).

Linear estimates for $\alpha$ can then be obtained through the following least squares specification (see also \cite{Gabaix2011}): 
\begin{equation}
    \label{EQ:regalpha}
    ln((i- 1/2) \bar{D}_i/\bar{D}) = C - \alpha  ln(w_{i})
\end{equation}

As discussed earlier, \citet{chakraborty2018missing} and \citet{Waltl2018} showed that this estimator produces unbiased and consistent estimates of $\alpha$ when information on top tail observations is provided.
The rich list sample is only used for the estimation of the Pareto tail parameters $\alpha$ and $w_{0}$. Afterward, the adjustment method is applied to survey sample $\mathcal{S}_0$.

The third step of the adjustment consists of estimating the total wealth in the top tail  $\hat{t}(w; top)$ by multiplying the total number of rich households $D$ resulting from the $S_0$ sample by the mean of the estimated Pareto distribution (given by $\alpha w_0 / (\alpha -1)$ for $\alpha > 1 $). We will later use this information to calibrate the sampling weights of rich households in the survey to the total wealth implied by the Pareto adjustment.   

This approach assumes that the sample estimate of $D$ (the total number of rich households) is unbiased. Indeed, some households have zero probability of being included in the survey (the missing tail from now on) after wealth reaches the truncation point $w_1$. This may be due to the difficulties in contacting such rich households to even negotiate an interview, or to a specific decision by the data producer to exclude them for operative or confidentiality reasons. Appended rich list observations will rarely be representative of all missing households. Also, the presence of differential non-response will imply that observed households in the Pareto tail are also under-represented as the probability of a household being interviewed approaches zero the closer its wealth is to the truncation point.

As a result of the underestimation of households in the Pareto tail, estimates for total wealth in the tail will also be underestimated. We then propose a novel method for the estimation of the number of missing rich households and their wealth.

Consider the sample $\mathcal{S}_T$ of Pareto-tailed households ordered by their wealth, and recall that $w_1$ is the truncation point above which there are not rich households in the sample. Following from the Glivenko-Cantelli theorem, because of the truncation the empirical cumulative distribution function resulting from this sample is different from the theoretical distribution implied by the Pareto adjustment. 

In particular, the following relation holds:
\begin{equation}
    \label{EQ:Glivenko-Cantelli}
    \inf_{i\in \mathbb{R}} \frac{D - D_{i-1}}{D} - \left(1- \left( \frac{w_{0}}{w_{i}}\right)^\alpha \right)  \geq 0
\end{equation}
where $D_{i-1}$ is the sum of weights of all households richer than $w_i$ ($D_i = \sum_{j=1}^{i}d_j$) and $D$ is the sum of the survey weights of observations in the survey Pareto tail (so that $D - D_{i-1} =\sum_{j=i}^{m}d_j$). This relation means that the empirical CDF will always suffer from a bias equal or larger than zero since units whose wealth exceeds $w_1$ are unobserved.

The theoretical Pareto CDF can then be used to correct the survey-based estimate by dividing the cumulative sum of survey weights for any point by the value of the Pareto CDF at that point:

\begin{equation}
    \label{EQ:nohh}
    t(d_i; top) \approx \frac{D - D_{i-1}}{1-(w_{0}/w_{i})^{\alpha}}.
\end{equation}

Analytically, the estimate from equation \eqref{EQ:nohh} should be the same for each $i$-th observation in the tail. In practice, with empirical data, variability in survey weights will affect the estimate of the number of households in the tail. Because of differential non-response, this becomes a particularly relevant problem when weight quality can deteriorate the closer observed wealth gets to the truncation point. The estimate can then be improved by estimating $t(d_i; top)$ for each value of wealth over a range of top tail observations, then estimating the mean $\hat{t}(d; top)$ as follows:

\begin{equation}
    \label{EQ:nohhhat}
    \hat{t}(d; top) = \frac{1}{m} \sum_{i=1}^m \frac{D - D_{i-1}}{1-(w_{0}/w_{i})^{\alpha}}
\end{equation}

An estimator of the number of missing, unobserved, households after the truncation point can be computed as $\hat{t}(d; miss) = \hat{t}(d; top) (w_{0}/w_{1})^{\alpha} $.  To account for these missing households, the total of observable households will be estimated as $\hat{t}(d; obs) = \hat{t}(d; top) (1 - (w_{0}/w_{1})^{\alpha})$. 

Finally, the total wealth in the top tail  $\hat{t}(w; top)$ can be estimated by the product of the estimated number of households and the Pareto mean:
\begin{equation}
    \label{EQ:totadj}
    \hat{t}(w;top) = \frac{\alpha w_0 }{(\alpha -1)}\hat{t}(d; top)
\end{equation}
Wealth in the missing part of the tail can similarly be computed as: $\hat{t}(w; miss) = \hat{t}(d; miss) \alpha w_1 / (\alpha -1)$, setting the new threshold at the truncation point $w_1$.\footnote{This is possible because the Pareto shape parameter does not change along the Pareto distribution.} 

\subsection{Calibration}
\label{subsection:SvyCalibration}

Calibration is a method whose aim is to correct the sampling weights 
$d_i$ through re-weighting methods while keeping the individual responses $w_i$ unchanged \citep{DevilleSarndal92, sarndal2007calibration}. In the literature, this approach is referred as design-based and it is mainly used: $(i)$ to force consistency of certain survey estimates to known population quantities; $(ii)$ to reduce non-sampling errors such as non-response errors and coverage errors; $(iii)$ to improve the precision of estimates \citep{haziza2017construction}.
 
Alternatively, the so-called model-based approach aims at adjusting the individual responses collected through the survey $w_i$ while sampling weights 
$d_i$ are left unchanged. It requires a model for the distribution of the measurement error and auxiliary information to estimate the parameters of the model. Among the several models available in the literature, those most suitable for our purposes are imputation methods. For a general description, see the seminal works by Rubin (\citeyear{Rubin76}, \citeyear{Rubin1987}). 

The two approaches have some shared traits, so that the distinction is not always clear-cut. For example, the weighting adjustment can also be seen as a method of imputation consisting of compensating for the missing responses by using those of the respondents with the most similar characteristics; in the same way, the imputation of plausible estimates in lieu of respondents’ claimed values can be thought of as a re-weighting method.

The choice of the method of adjustment is driven by three factors. First, it depends on the estimator of interest. For example, if the interest is to estimate the share of total wealth held by rich households, the use of the Pareto method (as described in section \ref{subsection:ParetoTail}) could be sufficient. Second, the choice depends on the magnitude of the gap to fill and the reasons behind it. If the gap is considerable and depends on both measurement error and non-response, one single approach may not be sufficient. Therefore, one may need to combine several methods. Finally, the choice depends on the information that is available. If, for example, the only available auxiliary information is in the form of population totals, then the calibration approach might be the only feasible way. However, if auxiliary data are available at the individual level, then the model-based methods may represent the most effective solution.


In the design-based approach, the calibration method for estimating the population total of a variable of interest is addressed through the following optimisation problem for finding a new set of weights $d^*_i$:
\begin{equation}
    \label{equation:calib}
    \min_{d^*_i} \sum_{i=1}^{S} c_iG(d_i^*;d_i)  \quad s.t. \quad  t(y)= \sum_{i=1}^{S}d_{i}^{*}y_{i},
\end{equation}
where $d_{i}^{*}=d_{i}a_{i}$, $c_iG(d_i^*;d_i)$ is a distance function between the basic design weights and the new calibrated weights, $c_i$ are known constants the role of which will be discussed in more detail later, and  $y$ represents an auxiliary variable, possibly vector valued. The adjustment factor $a_i$ is a function of the value on the sample of the variables used in the calibration procedure $y_i = (y_{i1}, y_{i2}, ..., y_{ik})$, and it is computed so that final weights meet benchmark constraints, $t(y)$, while, at the same time, being kept as close as possible to the initial ones. Closeness  can be defined by means of several distance functions \citep[see table 1 in][]{DevilleSarndal92}, the most common being the chi-squared type 
\begin{equation}
    \label{equation:chidist}
c_iG(d_i^*;d_i)=\frac{(d^*_{i}-d_{i})^2}{d_{i}c_i}
\end{equation}
for which an analytical solution always exists. The benchmark constraints are defined with respect to $t(y) = (t(y_1), t(y_2), ..., t(y_k))$, that is the known vector of population totals or counts of the calibration variables.

The final output is a single new set of weights to be used for all variables. The magnitude of the adjustment factors and therefore the variability of the final set of weights is a function of the number of constraints (dimension $k$ of the vector $t(y)$) and the imbalance (the difference between the Horvitz-Thompson estimate and the population total). Very variable weights hinder the quality of final estimates for sub-populations and for variables that are not involved in the calibration procedure. For these reasons, weights are usually required to meet range restrictions such as to be positive and/or within a chosen range. This can be achieved by suitably choosing and tuning the distance function $G(\cdot)$. 

The method was originally proposed to improve the efficiency of the estimators and to ensure coherence with population information, but then it was also largely applied to adjust for non-response \citep{Sarndal2005}. For example, \citet{little2005does} showed that if the variables used to construct the weights are associated both with non-participation and with the variable of interest, the bias and the variance of the estimator are reduced.

The main problem with the use of household balance sheet data in re-weighting methods is that wealth is generally skewed and concentrated in the hands of a small group of the population that has both low propensity to participate in the survey and different socio-demographic characteristics from the average population. 

\subsection{Adjusting for non-response: Pareto-calibration} \label{subsection:parcal}

We begin by exploiting the information obtained after fitting a Pareto distribution, as in subsection \ref{subsection:ParetoTail}, to adjust the wealth distribution in the survey for differential non-response using the calibration methods described in section \ref{subsection:SvyCalibration}. 

We proceed by using $\hat{w}_{0}$ and $\hat{\alpha}$ and equation \eqref{EQ:nohhhat} to estimate the total number of observable households over the threshold $\hat{t}(w; obs)$ and their total net wealth $\hat{t}(w; obs)$ (and the corresponding figures for households below the threshold ${w}_{0}$).

We then calibrate the sampling weights from sample $\mathcal{S}_0$ using the following constraints:  

\begin{equation}
    \label{EQ:calibvector}
    t(y)_1 = (\hat{t}(w; obs), \hat{t}(d; obs), t(w; bot), \hat{t}(d; bot), t(x))
\end{equation}

where $\hat{t}(d; obs)$ is the estimated number of observed households in the Pareto tail, $\hat{t}(d; bot)$  relates to the observations not in the tail, $\hat{t}(w, obs)$ is the estimated observable wealth in the Pareto tail, $t(w; bot)$ is a vector of Horvitz-Thompson estimators decomposing the initial wealth of observations below the threshold into their corresponding portfolio items,\footnote{Calibrating weights in the bottom part of the distribution to the initial, unadjusted, wealth in that part of the survey, average wealth among these observations will increase. To account for this issue, the calibration benchmark could be adjusted by subtracting $t(w; bot) - (t(d; bot)-\hat{t}(d; bot))t(w; bot)/t(d; bot)$. However, the disparity between the number of households in the Pareto tail and the ones in the bottom part of the distribution is so large that this adjustment is unlikely to affect our analysis. Therefore, in order not to over-stress the computational requirements of the model and focus on the part of the distribution where the effect of Pareto-calibration is significant, wealth in the non-Pareto part of the survey has been kept fixed.} and $t(x)$ is a vector of population counts for demographic characteristics.

Let the indicator variable $I_i = 1$ for $w_i \geq w_0$ and $I_i=0$ otherwise, then set the auxiliary variables vector for calibration to
\begin{equation}
\label{EQ:auxiliaryvector}
    {y}_{i1} = (w _i I_i, I_i ,  w_i (1-I_i), 1-I_i ,  x_{i})
\end{equation}

After calibrating survey data to these parameters, we obtain non-response adjusted weights $d^*$. This approach will be referred as `Pareto-calibration' from now on.

Should the survey be suffering from differential non-response issues only, this step might be sufficient to fill the gap with financial accounts. However, this is not always the case: provided that the we have a good approximation of wealth distribution in the tail, the remaining differences in coverage between the estimate obtained in equation \eqref{EQ:totadj} and the national accounts will then be left to measurement error.

\subsection{Adjusting for non-response and measurement error: Simultaneous approach} \label{subsection:simultaneous}

In order to correct for measurement error, we combine the adjustment for differential non-response described in subsection \ref{subsection:parcal} with the following procedure.

The first step is to run the Pareto-calibration adjustment, as described earlier. Let $d^*_i$ be the final weight from the non-response adjustment procedure.
 
As second step we run a  calibration procedure as in \eqref{equation:calib}  in which $(i)$ the $d^*_i$'s are considered to be the basic weights and $(ii)$ the set of benchmark constraints $t(y)_2$ are given by the macro aggregates. 
The adjustment factor $a_i$, for $i=1,\ldots,S$, obtained by this procedure is such that
\begin{equation}\label{eq:ai1}
\sum_{i=1}^{S}d_{i}^{*}a_iy_{i}=t(y)_2
\end{equation}
We apply this adjustment factor directly to the variables of interest so that 
\begin{equation}\label{eq:ai2}
y_{i}^*=a_iy_{i}.
\end{equation}
This approach shares similar traits with reverse calibration introduced by \citet{chambers2004outlier} to deal with outlier-robust imputation. 

Recall that $y_i$ is vector-valued. Then, note that this calibration is multivariate because it accounts for all constraints with respect to macro estimates in a single procedure and, therefore, it accounts for the multivariate structure of the variables included in $y$. In addition, every household has a different adjustment factor $a_i$ that depends on all the values of $y$. 

A special case of multivariate calibration is proportional allocation, which consists of allocating the  gap by multiplying each component of $y_i$ by the corresponding inverse of the item-specific coverage ratio.\footnote{In fact, if we focus on a single item, $y_1$, the adjustment factor used by proportional allocation can be obtained as the solution to a univariate calibration procedure in which $(i)$ the starting weights are again the $d_i^*$'s, $(ii)$ there is only one benchmark constraint $\sum_{i=1}^{S}d_{i}^{*}a_iy_{1i}=t(y_1)$, and $(iii)$ the distance function $G(\cdot)$ is chi-squared as in \eqref{equation:chidist} with constants $c_i=1/y_{1i}$. The proof is omitted for brevity, but it is close in spirit to Example 1 in \citet{DevilleSarndal92}}

This equivalence sheds some light on the role of the constants $c_i$'s in the distance function (\eqref{equation:calib}). In univariate calibration, if they are chosen to be the inverse of the variable in the constraint, then the adjustment factors are shrunk towards a common value for all households as in proportional allocation. On the contrary, if they are set to be constant, the adjustment factors would be roughly proportional to the values of the item. For this reason, in the proposed multivariate calibration for imputation, we have set the constants to possibly depend on the wealth of the household, that is  
\begin{equation}
    \label{eq:tau}
    c_i=\left(\frac{1}{w_i}\right)^\tau,
\end{equation}
where $\tau\geq 0$ can be seen as a shrinkage factor: larger values provide  adjustment factors that are more uniform across households, while values towards 0 provide adjustment factors with a higher variability and correlation with $w_i$.\footnote{For this work, we set $\tau = 1$. Future research might seek to retrieve information on $\tau$ using external data where no misreporting behaviour is present.}

In order to account for the missing wealthy households, we add a single observation with weight $\hat{t}(d;miss)$ and wealth $\hat{t}(w;miss) / \hat{t}(d;miss)$ is created and imputed at the top of the sample. This observation's portfolio is also allocated using portfolio shares in the Pareto tail of the distribution.

At the end of the multivariate calibration the gap is filled.
However, the distribution of $w_i$ has changed, because its components have changed. Some households which were initially classified as not rich may have moved in the top tail of wealth distribution. Therefore, we need to find the new Pareto threshold, and apply again the Pareto-calibration procedure described earlier. This requires an iterative procedure that alternates a Pareto-calibration step that  improves coverage and a multivariate calibration step that addresses measurement error. The two steps are iterated until convergence. Convergence has been set on the parameter $\alpha$ of the Pareto distribution: if the estimated values in two consecutive steps differ by less than a small predefined threshold the procedure stops \footnote{It is worth stressing that the converge of the process could also not be achieved, especially in the case the gap to be filled is sizeable.}.  

\subsection{A special case: Single-iteration approach}
\label{subsection:wealthcalib}

If one is willing to assume that (1) that relative error is independent from the observed wealth, at least among the very rich, and that (2) the relative error converges in probability to a constant, which we will denote $(w^* - w)/w \overset{p}{\to} \zeta$, so that, on average, the unobserved `true' total wealth will be given by $\hat{w}_{i}^* = \zeta w_{i} $, provided that $\zeta \perp w$, the method simplifies.

Thanks to Slutsky's theorem, survey wealth would still be Pareto distributed with tail parameter $\alpha$ after adjusting for measurement error. As it follows, total wealth in the survey would scale up to $\sum_{i=1}^{S} \zeta d^*_{i} w_{i} $, and  
the Pareto CDF would turn into $F_{\alpha}(\zeta w_{i})= 1 - (\zeta w_{0} / \zeta w_{i})^{\alpha}$.

Simplifying this last formula and updating equation \eqref{EQ:totadj} for measurement error, we obtain the following estimate for total wealth:

\begin{equation}
    \label{EQ:totadj_underrep}
    \zeta \hat{t}(w) =\zeta( \frac{\alpha w_0}{(\alpha -1)}\hat{t}(d; top) + \sum_{i=s}^{S}d_{i}^*w_{i} )
\end{equation}

This means that our estimate for $\alpha$ does not depend on the scaling of the variables. In this case, the coefficient for the Pareto-adjusted coverage ratio, given the national accounts total wealth, as in $\zeta = t(w) / \hat{t}(w)$, will yield the scalar to which to re-allocate reported survey wealth. It is straightforward that, to account for the missing wealth, wealth should be scaled to $\zeta (\hat{t}(w) - \hat{t}(w, miss))$, which, after Pareto-calibration, simplifies to $\zeta \sum_{i=1}^{S}d_{i}^*w_{i}$. 

As the Pareto shape parameter is unaffected by the re-scaling, the iterative procedure would no longer be needed. The adjustment for measurement error and for non-response at the tail of the distribution can be run independently from each other. 

In theory, because of the assumptions above mentioned, whatever the adjustment method for measurement error is used, the final data should still be Pareto distributed among rich households.

In practice, if one wants to make sure that this is the case, it is advisable to correct for measurement error using calibration in a slightly different manner than the one described in section \ref{subsection:SvyCalibration}.
Traditional calibration methods find the optimal adjustment factor $a_i$ which minimises the quadratic distortion of new weights relative to prior ones. We propose to change the objective function so that the adjustment factor $a_i$ is minimised with respect to a quadratic loss function for reported wealth values, as follows:

\begin{equation}
    \label{equation:calibwealth}
    min \sum_{i=1}^{S} \frac{(a_{i}w_{i}-w_{i})^2}{w_{i}} \quad s.t. \quad  \zeta \sum_{i=1}^{S}d_{i}w_{i}= \sum_{i=1}^{S}d^*_{i}a_{i}w_{i}
\end{equation}

In this method, the correction for measurement error is based on univariate calibration using total wealth as a sole benchmark. As the objective function minimised distortions relative to the initial reported value, the final imputed data will be Pareto distributed. 

\section{Assessment of the method}
\label{section:assessment}
The ideal approach for assessing the quality of the results would be to compare them with an external benchmark, for instance, coming from highly reliable administrative records. 
Without such auxiliary information, we can assess the method in two ways.
First, we assess the robustness of our results by comparing them with  other estimators based on different assumptions. Second, we assess the precision of our results  by estimating their variability .  

Beyond our simultaneous approach, we compute five alternative estimators:
\begin{itemize}
    \item ‘Survey \& missing tail’. The results are produced using the unadjusted survey data, plus an estimation of the total wealth held by rich household with zero probability of being in the survey (missing tail).
    \item ‘Pareto-calibration \& missing tail’. Survey data are adjusted with the Pareto-calibration model. Survey weights are calibrated and the total wealth of the missing tail is included in the estimate.
    \item `Pareto-calibration, proportional allocation \& missing tail’. This method adds to the previous one a correction for measurement error based on proportional allocation, as in \citet{oecd2013}. This is a very naive method based on the assumption that measurement error is equal across households and that it only depends on the instrument. Moreover, it does not enable to adjust for no-reporting.
    \item ‘Single-iteration approach \& missing tail’. In this method, the correction for measurement error is based on univariate calibration method described in subsection \ref{subsection:wealthcalib}. 
    Adjustments are applied on the y variable (gross wealth). After rescaling the threshold $w_0$ to account for measurement error, the missing tail is re-estimated and included.
    \item ‘Single-iteration approach, portfolio calibration \& missing tail’. This method extends the previous one by adding an extra step in which portfolios are calibrated using financial accounts totals -- adjusted to account for the missing part of the tail -- and Pareto distributional information as benchmarks. Calibration in the extra step works again on weights.
\end{itemize}

Variance estimation in our methodology has two main components. The first one is the sampling variance, which indicates the variability introduced by choosing a sample instead of enumerating the whole population, assuming that the information collected in the survey is otherwise exactly correct. 
A second source of variability is imputation variance which refers to the fact the methodology for filling the gap can produce several different plausible imputed data sets. The uncertainty due to the imputation process adds up to the sampling variance.

To estimate the overall variability we use the Rao-Wu rescaled bootstrap weights released with HFCS data to account for sampling variability \citep{hfcs2020a}.
For each of the 1,000 sets of bootstrap weights
we replicate all the methods previously described. In each replication, the parameters of the Pareto distribution are re-estimated introducing additional variability. We then obtain 
the mean and standard deviation from all successful simulations\footnote{A simulation is flagged as unsuccessful, and discarded, whenever a calibration procedure fails because of lack of convergence under the chosen restraints.} and compute the coefficient of variation to evaluate the robustness of our methods and derive a measure of their variability.

\section{Application to the HFCS}
\label{section:application}

\begin{figure}[H]
\includegraphics{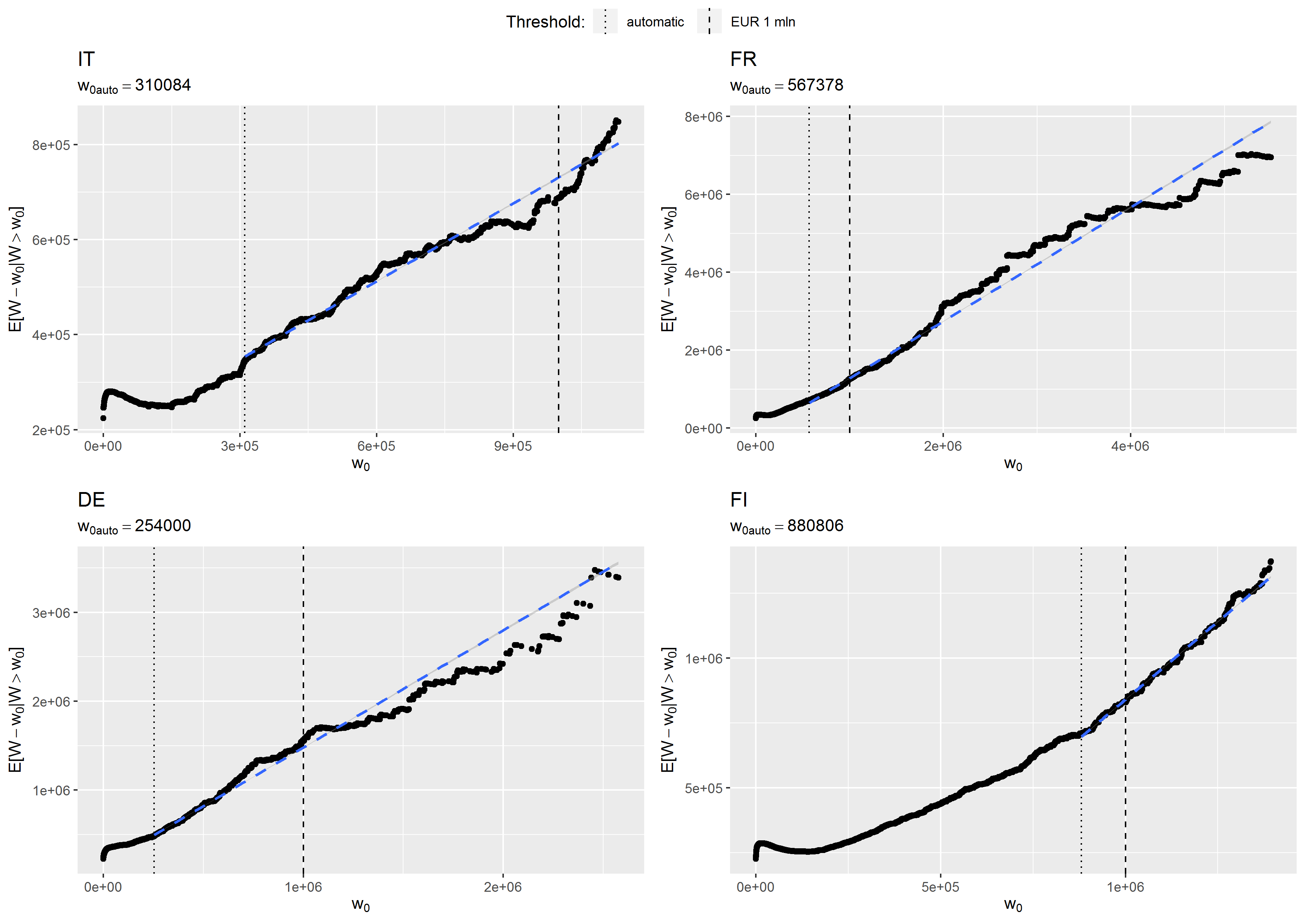}
\caption{\title{Pareto Threshold detection.} Mean excess plots for gross recorded wealth in the HFCS. Predicted Pareto thresholds and linear fits estimated using the proposed methodology.}
\label{figure:Parw0}
\end{figure}

The method described in the previous sections has been applied to the second 2014 wave of the HFCS.
The first step consists of estimating the parameters of the Pareto distribution ($w_{0}$ and $\alpha$). Figure \ref{figure:Parw0} provides a graphical intuition of the automatic selection of threshold for the four selected countries, showing the estimated $w_{0}$ and showing, given this threshold, linear fits for the mean excess conditional on wealth. Table \ref{tabAlphas} summarises the final results. As it appears, this approach provides benefits over an arbitrary threshold selection: in all cases, the new threshold is found to be lower than €1 million, meaning that subsequent estimates on tail behaviour will significantly benefit in precision.

Figure \ref{figure:ParDemoLin} illustrates the outcome of the Pareto-calibration process, showing the empirical CCDF on a log-log scale before and after the adjustment.  
Re-weighted figures are produced by using the proposed Pareto-calibration method; $\alpha$ indicates the Pareto shape parameter estimated by imputing the rich list, while $\theta$ shows these estimation results with survey data only.

Table \ref{tabCRatios} shows coverage ratios between survey wealth estimates and financial accounts. Column (1) shows initial coverage ratios, while column (2) displays coverage ratio for adjusted data, and column (3) grosses up survey wealth by estimating total wealth after truncation and adding it to the previous estimate. Columns (4) and (5) show the estimated number of households in the Pareto tail, along with the number of ``missing rich''. 

Overall, these figures suggest that the proposed Pareto-calibration approach can produce substantial improvements in survey coverage, especially in the absence of over-sampling or administrative data.  In the case of Finland and Germany, the discrepancies between micro and macro figures virtually disappear after calibrating survey data and accounting for the unobservable households. Coverage is also significantly improved for Italy and France, but the persistence of a mismatch between survey data and financial accounts points to the presence of measurement error.

Having re-estimated the number of households in the Pareto tail of the survey, our method also shows substantial improvements in coverage over the grossing up methods already explored in the literature, and suggests that adjustments for non-response should also focus on correcting the number of households in the Pareto tail, rather than only the wealth contained in it.

After dealing with the issue of nonresponse at the tail of the distribution, we use multivariate calibration to adjust for measurement error along the whole distribution. 

As benchmark constraints $t(y)$ we use the financial instruments with high conceptual comparability between survey and financial accounts -- namely, deposits, bonds, shares, funds, insurances and pensions, money owed to the household and liabilities -- following from the comparability scale provided by \citet{eglmm}. The resulting adjustment factors are then applied to financial instruments with lower comparability -- business and housing wealth -- which, assuming that measurement error is comparable within comparable financial instruments, should ensure that the adjustment will not be biased by the presence of instruments with low comparability.

We then iterate the Pareto-calibration and the multivariate calibration until convergence. Convergence has been set on the parameter $\alpha$ of the Pareto distribution: if the estimated value in two consecutive steps differs by less than a small predefined threshold,\footnote{In the current application, this tolerance was set at 0.05.} the procedure stops. Convergence is usually achieved in a limited number of steps (between 1 and 3 in the application at hand). 

Table \ref{tabGfactor} shows the average values of the adjustment factors $a_i$'s (as well as 
coefficients of variation) as a function of gross wealth percentiles at the end of the iterative procedure for the four countries. That is, these are the overall adjustment of the survey variables at the end of the procedure obtained as the ratio between the final imputed values and the ones from the original survey. 

\section{Results}
\label{section:results}

Table \ref{tabTopBoot} shows distributional results indicating the proportion of net wealth held by the top 1, 5, 10, and 20 weighted percentiles, along with the bottom 50\%. Weighted Gini inequality indices are also presented in column (6), while column (7) provides the estimated Pareto tail parameter $\alpha$ given the data. These figures are reproduced under each allocation method. The bootstrap-based coefficient of variation is reported in parentheses for each estimate. 

The first set of rows (`Base Survey') presents distributional figures from the unadjusted HFCS data. As is well known, truncation in top wealth distribution and measurement error can cause survey estimates to understate the true level of wealth inequality, and the figures presented in the table provide support for this possibility. Indeed, estimates from the unadjusted HFCS would suggest wealth inequality in Italy, which has one of the largest micro-macro gaps, to be close to the inequality level in Finland, where the gap is lower.

Column (7) displays the Pareto tail coefficients. In the first set of rows, the $\alpha$ parameter is estimated using survey data only, meaning that this is the Pareto estimate that survey data yields when truncation is not corrected through the imputation of a rich list.

For all following sets of rows, which correspond to the alternative estimators discussed in section \ref{section:assessment}, we also include an adjustment for the unobserved part of the Pareto tail as presented in section \ref{subsection:ParetoTail}. To do so, these missing households are imputed as a single observation in which the weight and wealth are respectively equal to the estimated number of unobserved households and the estimated average wealth in the unobserved Pareto tail.

The second set of rows (`Survey \& missing tail') displays estimates produced using the un-adjusted survey data, plus the missing tail households. Depending on the size of the truncation in the Pareto tail, inequality estimates can be affected considerably. For surveys, such as the Italian and German ones, in which truncation bias is particularly pronounced, the sole inclusion of these unobserved households increases the proportion of wealth held by the top 1\% households by at least 10.7 and 10.3 percentage points, respectively. This increase is much less pronounced for the French and Finnish surveys, where the truncation is also much more modest.

\begin{figure}[H]
\includegraphics{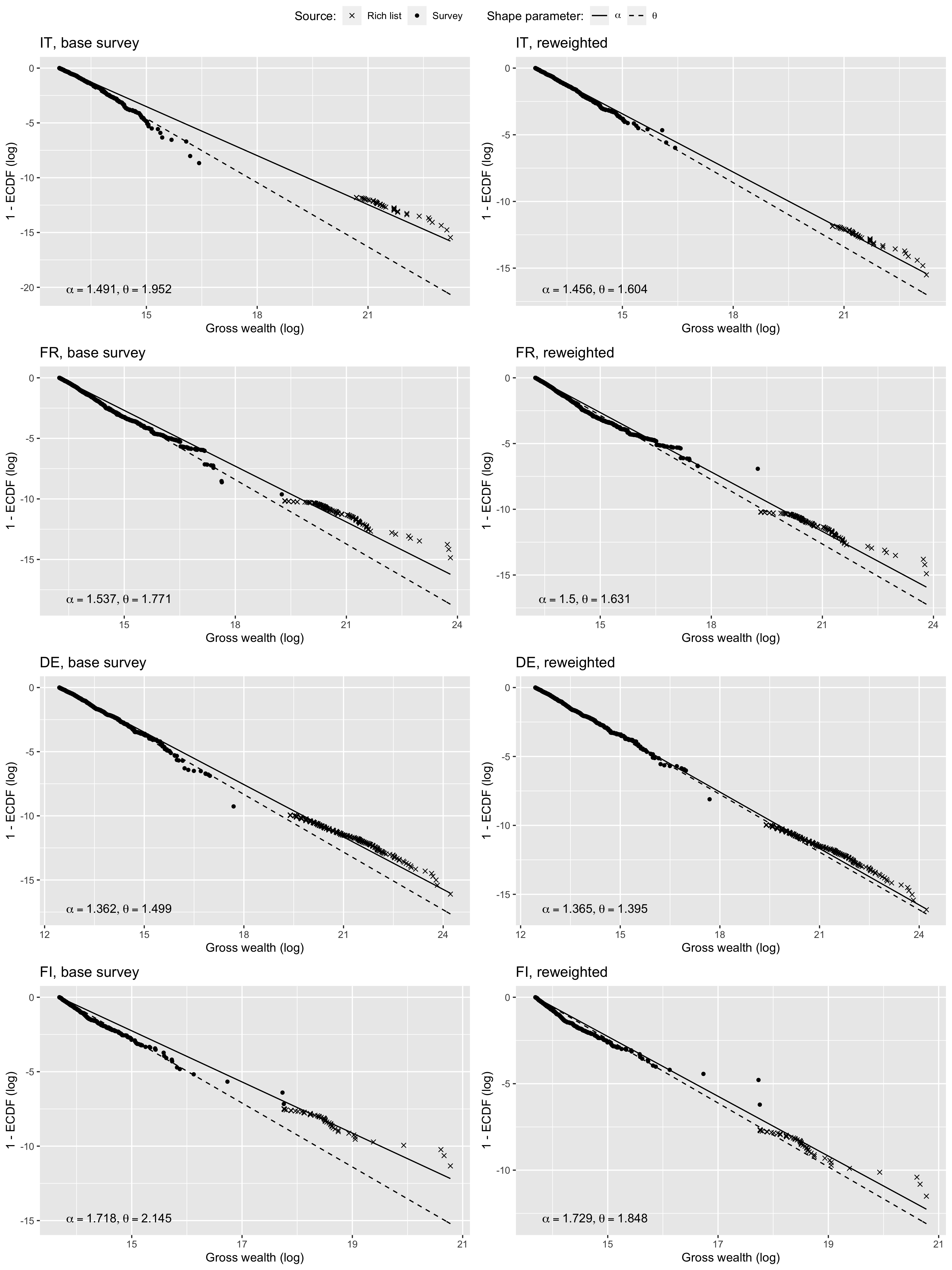}
\caption{\title{Pareto Tail Re-weighting.} Empirical cumulative distribution functions (log scale) for survey wealth distributions in the Pareto Tail. Re-weighting achieved by using the Pareto-calibration method, using the calibration benchmarks from equation \eqref{EQ:calibvector}. $\theta$ parameters estimated using survey data only, $\alpha$ estimated using Vermeulen's \citeyear{vermeulen2018fat} regression method with imputed rich list.}
\label{figure:ParDemoLin}
\end{figure}

The inclusion of the unobserved tail raises inequality levels for all the surveys considered, but again these increases are proportional to the size of the truncation. Finally, estimates for the Pareto tail parameter are now corrected for the truncation by imputing the rich list and using the estimation procedure described in section \ref{section:Methodology}. These are the same parameters earlier shown in Figure \ref{figure:ParDemoLin}.

Survey weights are then adjusted using the proposed Pareto-calibration method to produce the figures shown in the third set of rows (`Pareto-calibration \& missing tail'). After this adjustment, between-country differences across distributional indicators start to decline. This time, an increase in inequality, while less remarkable than in the previous step, can still be noted across all surveys.  Should there be a reason to suspect that survey weights degrade due to differential non-response, this increase suggests that the proposed adjustment can make an important contribution in the measurement of inequality through the adjustment of existing survey data points. 

After applying the adjustment, the tail parameters are re-estimated and shown in column (7). Their closeness to the initial Pareto estimates, shown in the previous set of rows, suggests that the calibration process does minimise distortions from the estimated Pareto distribution, even in cases in which the issues of truncation and differential-non-response are more severe. Improvements can be noted over the $\theta$ parameters (obtained without imputation of the rich list) as well, which are now closer to the rich-list imputed $\alpha$ parameters, as shown in figure \ref{figure:ParDemoLin}.

The row sets from fourth to sixth adjust the survey applying the estimators described in section \ref{section:assessment}. For countries like Finland and Germany, where measurement error seems to be a negligible issue, these adjustments might not be needed, and remaining divergences in portfolio item coverage against macroeconomic aggregates should be treated as sampling issues and adjusted through weight calibration, as detailed in section \ref{section:assessment}, and shown in the last set of rows in table \ref{tabTopBoot}.

In the fourth set of rows (`Par-cal, proportional allocation \& missing tail'), portfolio items are scaled proportionally to the Financial Accounts aggregates. Proportional allocation, however, seems like an inadequate solution. While proportionally allocated items do not generate severe distortions in the estimated Pareto distribution, the proportional allocation will most likely affect the portfolio allocation within each household. Since it is based on very unreliable assumptions this method should be considered in cases where the gap to fill is minimal. 

The fourth (`Par-cal, Wealth calibration, \& missing tail') and fifth (`Par-cal, Wealth/portfolio calibration, \& missing tail') sets of rows show how distributional figures are affected by the approach described in subsection \ref{subsection:SvyCalibration} .

In both cases, substantial differences over proportional allocation can be noted. First, the Pareto tail parameter is always closer to the initial estimate, meaning that the reallocation process, this time, leaves the distributional features of the survey intact. Secondly, inequality figures appear to be much more like the estimates produced in the previous steps. Indeed, the final output shows comparable results across all surveys, in which the increases in inequality, compared to the initial survey data, are proportional to the severity of both truncation and measurement error problems.

Most importantly, the $\alpha$ parameter is still close enough to the one estimated initially, suggesting, once again, that neither adjustment gives rise to unnecessary distortions in the tail wealth distribution. This is a relevant result, that validates the assumption of relative error converging in probability to a constant.

While wealth calibration should not be treated as a substitute for proper models for adjusting for measurement error, especially when this error is linked to socio-economic or behavioural factors, these calibration-based methods can still assist in the production of distributional figures without exposing the researcher to the risk of misrepresenting the distribution of household wealth and individual asset compositions.  

Also, the use of portfolio calibration (as in the penultimate set of rows) can help when measurement error is supposed to be null (Finland, and Germany to a lesser degree), and when models have been used to address such a problem. In these cases, the wealth calibration step can be skipped entirely, while the portfolio calibration can be paired with Pareto-calibration within the same step, so that the weighted sum of each portfolio item is kept consistent with the corresponding macro-economic aggregate, producing consistent and correct distributional figures.

The results obtained using the simultaneous approach are presented in the final set of rows. Here, we see that the distributional estimates are broadly in line with the results produced by the other methods. 
In particular, it provides very similar results to the Single-iteration approach, suggesting that its simplifying assumptions are likely to hold, at least in the four countries used in the analysis.    

Overall, all the methods consistently show that the household finance survey under-estimate the levels of wealth inequality. Moreover, the larger the wealth gap between micro and macro data, the higher the increase in the measures of inequality. 

As to variance estimation, the adjustment methods generally produce a decrease in the reliability of the results. This is expected since they add some additional variability because of the imputation process. 

For each method, the precision increases when the statistic relates the bottom or median part of the wealth distribution. The estimators of the wealth share held by the top 1 percent have a low precision in all countries. 

Compared to other methods, the simultaneous approach produces the lowest increase in variability. This is also due to the use of multivariate calibration, a method that has been originally developed to increase the precision of estimators. The final coefficients of variation are not very different from those based on the unadjusted survey data, especially for the statistics that do not relate to the top tail of the distribution.
\section{Conclusions}
\label{section:conclusions}

In this paper, we show how a combination of well-established methodologies for the fitting of a Pareto distribution and the calibration of survey data can be used to correct for non-response and misreporting when only limited external information is available. 

We apply these methods to the HFCS data, using the 2014 Finnish, French, German, and Italian surveys, and employing rich list data from Forbes or national press sources, along with household sector aggregates from national accounts, as auxiliary sources of information.

We show that these adjustment methods improve the production of distributional national accounts for the household sector, since inequality estimates from the survey data understate the population parameters, depending on the severity of both non-response and measurement error. 
We also discuss how to assess the quality of these distributional indicators. 

Further work is needed for the refinement of the methodology we propose. For example, the estimation of the number of wealthy households could be further validated and improved, for instance by using alternatives to rich lists (such as tax records) or by applying additional methods (such as the Type II Pareto or the Estate Multiplier Method). Also, the correction of measurement error could be further improved by enriching the auxiliary variables vector with more granular external information (if available).

Nonetheless, our framework has the advantage to offer a set of adaptable tools that can be fine-turned on a case-by-case basis. Indeed, both the Pareto-calibration adjustment and the multivariate calibration methods can be enhanced with external information and can be run separately when needed. 

Moreover, our contribution allows to compute several distributional indicators using the adjusted micro data-set, while most studies in the current literature only focus on providing aggregate estimates for wealthy households. 

\section*{Acknowledgements}

The paper has greatly benefited from the discussions with all members of the EG-LMM and EG-DFA working groups.

\bibliographystyle{apalike}
\bibliography{References.bib}

\newpage
\section{Appendix: Tables and figures}

\input tableAlphas.tex

\input tableCratios.tex

\input tableGfactors.tex

\newpage

\input tableTopsharesBoot.tex




\end{document}

%% file: tableAlphas.tex
\begin{table}[bht!]
\caption{The missing gap: Pareto tail parameters and estimated thresholds \label{tabAlphas}}
\begin{center}
\begin{tabular}{lccc} \toprule

 & $\theta$ & $\alpha$  & $w_0$ \\
Country & (1) & (2) & (3)  \\ 
\midrule

IT & 1.952 & 1.491 & 310,084 \\
FR & 1.771 & 1.537 & 567,378 \\
DE & 1.499 & 1.362 & 254,000 \\
FI & 2.145 & 1.718 & 880,806 \\
\bottomrule

\end{tabular}
\end{center}

\begin{tabular}{p{\textwidth}}
\small \textit{Notes:} Pareto tail parameters and estimated thresholds. $\theta$ parameters estimated using survey data only, $\alpha$ estimated using Vermeulen's \citeyear{vermeulen2018fat} regression method with imputed rich list. 
\end{tabular}

\end{table}

%% file: tableCratios.tex
\begin{table}[H]
\caption{The missing gap: Pareto adjustments for gross wealth \label{tabCRatios}}
\begin{center}
\begin{tabular}{lcccccc} \toprule

 & \multicolumn{3}{c}{Coverage Ratios} & \multicolumn{3}{c}{Estimated tail households}\\
 & Base & Adjusted & Implied & Total & 95\% C.I. & Missing\\
Country & (1) & (2) & (3) & (4) & (5) & (6) \\ 
\midrule

IT & 0.553 & 0.647 & 0.727 & 5,483,837 & $\pm$ 243.406 & 19366.110 \\
FR & 0.673 & 0.762 & 0.779 & 3,003,389 & $\pm$ 60.255 & 289.339 \\
DE & 0.827 & 0.908 & 1.039 & 9,889,923 & $\pm$463.185 & 7913.568 \\
FI & 0.917 & 1.019 & 1.037 & 108,507 & $\pm$36.287 & 91.788 \\
\bottomrule

\end{tabular}
\end{center}

\begin{tabular}{p{\textwidth}}
\small \textit{Notes:} Coverage ratios and estimated number of households in the tail. Re-weighting achieved with the Pareto-calibration method, using the calibration benchmarks from equation \ref{EQ:calibvector}.
\end{tabular}

\end{table}





%% file: tableGfactors.tex
\begin{table}[H]
\caption{Simultaneous approach: final multivariate calibration adjustment factors \label{tabGfactor}}
\begin{center}
\begin{tabular}{lS[table-format=1.3,
                    round-mode=places,
                    round-precision=3, table-number-alignment = center, table-column-width = 0.056\textwidth] S[table-format=1.3,
                    round-mode=places,
                    round-precision=3, table-number-alignment = center, table-column-width = 0.056\textwidth] S[table-format=1.3,
                    round-mode=places,
                    round-precision=3, table-number-alignment = center, table-column-width = 0.056\textwidth] S[table-format=1.3,
                    round-mode=places,
                    round-precision=3, table-number-alignment = center, table-column-width = 0.056\textwidth] S[table-format=1.3,
                    round-mode=places,
                    round-precision=3, table-number-alignment = center, table-column-width = 0.056\textwidth] S[table-format=1.3,
                    round-mode=places,
                    round-precision=3, table-number-alignment = center, table-column-width = 0.056\textwidth]
                    S[table-format=1.3,
                    round-mode=places,
                    round-precision=3, table-number-alignment = center, table-column-width = 0.056\textwidth] S[table-format=1.3,
                    round-mode=places,
                    round-precision=3, table-number-alignment = center, table-column-width = 0.056\textwidth] S[table-format=1.3,
                    round-mode=places,
                    round-precision=3, table-number-alignment = center, table-column-width = 0.056\textwidth] S[table-format=1.3,
                    round-mode=places,
                    round-precision=3, table-number-alignment = center, table-column-width = 0.056\textwidth] S[table-format=1.3,
                    round-mode=places,
                    round-precision=3, table-number-alignment = center, table-column-width = 0.056\textwidth] S[table-format=1.3,
                    round-mode=places,
                    round-precision=3, table-number-alignment = center, table-column-width = 0.056\textwidth]
                    } \toprule

 & \multicolumn{10}{c}{Percentile}\\
 & {0.10} & {0.20} & {0.30} & {0.40} & {0.50} & {0.60} & {0.70} & {0.80} & {0.90} & {1.00} \\
Country  & {(1)} & {(2)} & {(3)} & {(4)} & {(5)} & {(6)} & {(7)} & {(8)} & {(9)} & {(10)}\\ 
\midrule

IT & 1.37026106783785 & 1.94440779863392 & 1.95227702414148 & 1.17050116941896 & 1.08991912100276 & 1.08513143028942 & 1.1194212030138 & 1.14095589716323 & 1.18677959156013 & 1.32643318666427 \\
 & (0.110100478) & (0.128386592) & (0.185638498) & (0.053985327) & (0.019116371) & (0.016473403) & (0.015599364) & (0.021732155) & (0.01627191) & (0.097799845) \\

FR & 1.03447924352059 & 1.57610743824402 & 2.3873434528838 & 1.75403539163449 & 1.2211060987705 & 1.12997667254321 & 1.12195596456841 & 1.1263329682173 & 1.1337170289553 & 1.23084270910749 \\
& (0.102114312) & (0.151260448) & (0.066618173) & (0.195827762) & (0.046845418) & (0.005779879) & (0.004145259) & (0.006719404) & (0.008334635) & (0.112678639) \\

DE & 1.1189058940686 & 1.2780136912077 & 1.55871252851926 & 1.71477184319106 & 1.59634852450513 & 1.33869270844707 & 1.17306248428225 & 1.12185249146972 & 1.11040076603235 & 1.09962970264282\\
 & (0.117598313) & (0.077011781) & (0.040544981) & (0.024094583) & (0.04290484) & (0.06090941) & (0.017420882) & (0.009559354) & (0.006931982) & (0.017240653) \\

FI &  0.943071184741476 & 1.08568515940912 & 1.35001933624595 & 1.10319276719353 & 1.04152778920335 & 1.02804184112081 & 1.03259520309763 & 1.03951978214609 & 1.04670343772154 & 1.0448674536859\\
 & (0.092597014) & (0.090366998) & (0.057036395) & (0.043957307) & (0.004562964) & (0.003244137) & (0.004485249) & (0.002541498) & (0.00319745) & (0.028715268) \\

\bottomrule

\end{tabular}
\end{center}

\begin{tabular}{p{\textwidth}}
\small \textit{Notes:} Mean and coefficient of variation of overall adjustment factors $a_i$, equations (\ref{eq:ai1}) and (\ref{eq:ai2}), from the multivariate calibration approach for imputation as a function of gross wealth percentiles.
\end{tabular}

\end{table}

%% file: tableTopsharesBoot.tex
\begin{center}
\begin{longtable}{lS[table-format=1.3,
                    round-mode=places,
                    round-precision=3, table-number-alignment = center, table-column-width = 0.07\textwidth] S[table-format=1.3,
                    round-mode=places,
                    round-precision=3, table-number-alignment = center, table-column-width = 0.07\textwidth] S[table-format=1.3,
                    round-mode=places,
                    round-precision=3, table-number-alignment = center, table-column-width = 0.07\textwidth] S[table-format=1.3,
                    round-mode=places,
                    round-precision=3, table-number-alignment = center, table-column-width = 0.07\textwidth] S[table-format=1.3,
                    round-mode=places,
                    round-precision=3, table-number-alignment = center, table-column-width = 0.07\textwidth] S[table-format=1.3,
                    round-mode=places,
                    round-precision=3, table-number-alignment = center, table-column-width = 0.07\textwidth]
                    S[table-format=1.3,
                    round-mode=places,
                    round-precision=3, table-number-alignment = center, table-column-width = 0.07\textwidth] S[table-format=1.3,
                    round-mode=places,
                    round-precision=3, table-number-alignment = center, table-column-width = 0.07\textwidth]} 
\caption{The missing gap: distributional wealth indicators  \label{tabTopBoot}} \\ \toprule

 & \multicolumn{8}{c}{Wealth Shares}\\
 & {Top 1\%} & {Top 5\%}  & {Top 10\%}  & {Top 20\%}  & {Bot 50\%} & {Gini} & {Tail $\alpha$} & {S.r.}  \\
Country & {(1)} & {(2)} & {(3)} & {(4)} & {(5)} & {(6)} & {(7)} & {(8)} \\ 
\midrule
\endhead
& \multicolumn{8}{c}{Base Survey} \\

IT & 0.1135414 & 0.2928937 & 0.4237249 & 0.5977503 &  0.1033786 & 0.5971817 & 1.951613 \\
& (00.074043924) & (0.032349375) & (0.020676699) & (0.012581805) & (0.038947602) & (0.01230447) & (0.074651195) & \\

FR & 0.1777410 & 0.3612573 & 0.4942968 & 0.6636282 & 0.06892689 & 0.6630178 & 1.77076 \\
& (0.089322942) & (0.036388704) & (0.021976953) & (0.011554199) & (0.039716146) & (0.011225063) & (0.052288883) & \\

DE & 0.2258016 & 0.4432907 & 0.5754625 & 0.7442643 & 0.03001578 & 0.7410518 & 1.499281 \\
& (0.111291626) & (0.049686226) & (0.030760529) & (0.015098774) & (0.072603913) & (0.013842171) & (0.159378921) & \\

FI & 0.120004015 & 0.28575242 & 0.41621747 & 0.59493453 & 0.1008288 & 0.5964584 & 2.145392 \\ 
& (0.050382835) & (0.018772414) & (0.011529657) & (0.006889984) & (0.025254053) & (0.007033199) & (0.068581422) & \\

& \multicolumn{8}{c}{Survey \& missing tail} \\

IT & 0.2207094 & 0.3794400 & 0.4948255 & 0.6477326 & 0.09074253 & 0.6469042 & 1.490836  & 0.999 \\
& (0.140352891) & (0.068805967) & (0.044394832) & (0.024330073) & (0.056430522) & (0.019044346) & (0.015357159) & \\

FR & 0.1980013 & 0.3772809 & 0.5069829 & 0.6720342 & 0.06719778 & 0.6714585 & 1.537246 & 1.000 \\
& (0.25414802) & (0.110739881) & (0.066960475) & (0.03445033) & (0.078570837) & (0.020877417) & (0.030471995) & \\

DE & 0.3291251 & 0.5197045 & 0.6337345 & 0.7793665 & 0.02597528 & 0.7763417 & 1.361727 & 0.971 \\
& (0.077118545) & (0.037726654) & (0.02430974) & (0.012291083) & (0.073533691) & (0.011802833) & (0.017136201) & \\

FI & 0.1364734 & 0.2989109 & 0.4269752 & 0.6023960 &  0.09894179  & 0.6039581 & 1.717602  & 1.000 \\
& (0.071817044) & (0.026250988) & (0.015714254) & (0.008674642) & (0.026563672) & (0.00876459) & (0.039229624) & \\

& \multicolumn{8}{c}{Pareto-calibration \& missing tail} \\

IT & 0.2655057 & 0.4309853 & 0.5403791 & 0.6816255 & 0.08393818 & 0.676523 & 1.456047  & 0.999 \\
& (0.087269404) & (0.043929358) & (0.028611299) & (0.014907495) & (0.045318048) & (0.013948789) & (0.015357159) & \\  

FR & 0.2738163 & 0.4375406 & 0.5556797 & 0.7049778 & 0.06097243 & 0.7035584 & 1.499868  & 1.000 \\
& (0.181652505) & (0.083733685) & (0.050389792) & (0.024677847) & (0.038990646) & (0.023211861) & (0.029178852) & \\

DE & 0.3600762 & 0.5478988 & 0.6566814 & 0.7926404 & 0.02471128 & 0.7886409 & 1.364833 & 0.971 \\
& (0.083589631) & (0.039739964) & (0.023516879) & (0.010000952) & (0.070521832) & (0.01039992) & (0.017136201) & \\

FI & 0.1921673 & 0.3503730 & 0.4718573 & 0.6349555 & 0.09130164 & 0.6343717 & 1.717602  & 1.000 \\
& (0.164869007) & (0.075492925) & (0.046222362) & (0.022306239) & (0.029968827) & (0.019608429) & (0.039229624) & \\ 

& \multicolumn{8}{c}{Par-cal, Proportional allocation \& missing tail} \\

IT & 0.2986025 & 0.4704124 & 0.5807633 & 0.7147173 & 0.07749535 & 0.7164887 & 1.427452 & 0.999 \\
& (0.10791799) & (0.054076974) & (0.035016892) & (0.018325945) & (0.048250771) & (0.017345783) & (0.009200605) & \\

FR & 0.2386463 & 0.4078761 & 0.5307819 & 0.6871827 & 0.07023237 & 0.6989841 & 1.557694  & 0.996 \\
& (0.18129983) & (0.085520974) & (0.052837983) & (0.027007022) & (0.048272211) & (0.025463006) & (0.032084263) & \\ 

DE & 0.3370375 & 0.5169412 & 0.6319117 & 0.7741009 & 0.03251645 & 0.7336069 & 1.375624 & 0.971 \\ 
& (0.092227307) & (0.043820285) & (0.025985923) & (0.011021125) & (0.070916238) & (0.011560138) & (0.016247937) & \\

FI & 0.1901221 & 0.3485809 & 0.4704004 & 0.6339484 & 0.09155352 & 0.633372 & 1.735287  & 1.000 \\
& (0.171791997) & (0.079373743) & (0.048710654) & (0.023626459) & (0.030878846) & (0.020948357) & (0.039096703) & \\ 

&\\
&\\

& \multicolumn{8}{c}{Par-cal, Wealth calibration \& missing tail} \\

IT & 0.3002861 & 0.4645724 & 0.5686188 & 0.7016403 & 0.07837705 & 0.6743792 & 1.442174  & 0.984 \\
& (0.225486224) & (0.117403392) & (0.077133026) & (0.042237522) & (0.102170295) & (0.041774632) & (0.009196682) & \\

FR & 0.2644996 & 0.4283579 & 0.5475698 & 0.6980228 & 0.06323017 & 0.714199  & 1.534188 & 0.988 \\
& (0.229959407) & (0.110774328) & (0.067829808) & (0.033365981) & (0.067739478) & (0.033046372) & (0.033996809) & \\

DE & 0.3568284 & 0.5485280 & 0.6583804 & 0.7946897 & 0.02429046 & 0.7607378 & 1.365316 & 0.97 \\
& (0.084869232) & (0.045522824) & (0.030377721) & (0.015828329) & (0.079638546) & (0.014616543) & (0.01611418) & \\

FI & 0.1881814 & 0.3486941 & 0.4715287 & 0.6360913 & 0.09035164 & 0.6350853 & 1.706971 & 1.000 \\
& (0.167844416) & (0.084166918) & (0.056439226) & (0.031979879) & (0.049493716) & (0.027975696) & (0.038684284
) & \\

& \multicolumn{7}{c}{Single-iteration approach \& missing tail} \\

IT & 0.3028315 & 0.4699631 & 0.5800269 & 0.7117514 & 0.07284551 & 0.70682 & 1.43162 & 0.983 \\
& (0.296798357) & (0.121360125) & (0.076636548) & (0.042065957) & (0.110417105) & (0.042278155) & (0.009143348) & \\

FR & 0.2442837 & 0.4219004 & 0.5433124 & 0.6951758 & 0.06422416 & 0.6930525 & 1.510868 & 0.987 \\
& (0.245881102) & (0.104212265) & (0.059169995) & (0.026051977) & (0.070896338) & (0.026391168) & (0.038278876) & \\

DE & 0.3623710 & 0.5487513 & 0.6579050 & 0.7939098 & 0.02478678 & 0.7894852 & 1.365239  & 0.942 \\  & (0.114989255) & (0.050432023) & (0.031934494) & (0.016388484) & (0.090627147) & (0.015674941) & (0.020903145) & \\

FI & 0.1424634 & 0.3118930 & 0.4394885 & 0.6129370 & 0.09952350 & 0.610366 & 1.778829  & 0.992 \\
& (0.109527528) & (0.043780021) & (0.028971436) & (0.01681636) & (0.031674222) & (0.0125746) & (0.02869323) & \\ 

& \multicolumn{7}{c}{Simultaneous approach} \\

IT & 0.299274833407283 & 0.482612014251461 & 0.595207135420408 & 0.729665691577544 & 0.0693698902724493 & 0.707566217281463 & 1.46462128167641 & 0.999 \\
& (0.080747521) & (0.038100521) & (0.023442986) & (0.013136038) & (0.055437696) & (0.013130281) & (0.005709876) & \\

FR & 0.286843425846978 & 0.454142739500061 & 0.568708037989891 & 0.71031217682572 & 0.0714555067498729 & 0.698281057105233 & 1.50196176761406 & 0.996 \\
& (0.148302721) & (0.065476277) & (0.040231525) & (0.020834759) & (0.044095909) & (0.018487718) & (0.018619528) & \\

DE & 0.34049280728372 & 0.524236685918064 & 0.635677894333435 & 0.77367928588928 & 0.0342466988289143 & 0.766857031286634 & 1.38785073033214 & 0.97 \\
& (0.09207948) & (0.042801025) & (0.025120159) & (0.011536941) & (0.068353716) & (0.012167132) & (0.018151911) & \\

FI & 0.173207758827924 & 0.336311172263858 & 0.460605596531566 & 0.628015237595531 & 0.0969096807261284 & 0.616664243468137 & 1.74836019878326 & 0.999 \\
& (0.114902346) & (0.048577984) & (0.033591862) & (0.020074588) & (0.049705703) & (0.017467699) & (0.016293128) & \\ 
\bottomrule
\end{longtable} \end{center}
\begin{tabular}{p{\textwidth}}
\small \textit{Notes:} Wealth share by percentile, Gini inequality coefficients and Pareto tail parameters for Italy, France, Germany and Finland, estimated using different adjustments for the HFCS data, and accounting for the unobserved part of the Pareto tail. Bootstrap-based coefficients of variation reported in parentheses. Success rates (S.r.) report the observed probability of convergence for calibration.
\end{tabular}

%% file: main.bbl
\begin{thebibliography}{}

\bibitem[Aban et~al., 2006]{Aban2006}
Aban, I.~B., Meerschaert, M.~M., and Panorska, A.~K. (2006).
\newblock Parameter estimation for the truncated pareto distribution.
\newblock {\em Journal of the American Statistical Association},
  101(473):270--277.

\bibitem[Alvaredo et~al., 2018]{Alvaredo2018}
Alvaredo, F., Atkinson, A.~B., and Morelli, S. (2018).
\newblock Top wealth shares in the {UK} over more than a century.
\newblock {\em Journal of Public Economics}, 162:26--47.

\bibitem[Alvaredo and Saez, 2009]{Alva2009}
Alvaredo, F. and Saez, E. (2009).
\newblock Income and wealth concentration in spain from a historical and fiscal
  perspective.
\newblock {\em Journal of the European Economic Association}, 7(5):1140--1167.

\bibitem[Ampudia et~al., 2016]{AMPUDIA2016}
Ampudia, M., van Vlokhoven, H., and Żochowski, D. (2016).
\newblock Financial fragility of euro area households.
\newblock {\em Journal of Financial Stability}, 27:250 -- 262.

\bibitem[Bach et~al., 2019]{Bach2019}
Bach, S., Thiemann, A., and Zucco, A. (2019).
\newblock Looking for the missing rich: tracing the top tail of the wealth
  distribution.
\newblock {\em International Tax and Public Finance}, 26(6):1234--1258.

\bibitem[Blanchet et~al., 2018]{blanchetetal2018}
Blanchet, T., Flores, I., and Morgan, M. (2018).
\newblock The weight of the rich: Improving surveys using tax data.
\newblock {\em {WID.world WORKING PAPER SERIES N° 2018/12}}.

\bibitem[Blanchet et~al., 2017]{blanchet2017generalized}
Blanchet, T., Fournier, J., and Piketty, T. (2017).
\newblock Generalized pareto curves: theory and applications.
\newblock {\em {WID.world WORKING PAPER SERIES N° 2017/3}}.

\bibitem[Casiraghi et~al., 2018]{cas}
Casiraghi, M., Gaiotti, E., Rodano, L., and Secchi, A. (2018).
\newblock {A “reverse Robin Hood”? The distributional implications of
  non-standard monetary policy for Italian households}.
\newblock {\em Journal of International Money and Finance}, 85(C):215--235.

\bibitem[Chakraborty et~al., 2019]{Chakraborty19}
Chakraborty, R., Kavonius, I., Perez-Duarte, S., and Vermeulen, P. (2019).
\newblock Is the top tail of the wealth distribution the missing link between
  the household finance and consumption survey and national accounts?
\newblock {\em Journal of official Statistics}, 35:31--65.

\bibitem[Chakraborty and Waltl, 2018]{chakraborty2018missing}
Chakraborty, R. and Waltl, S.~R. (2018).
\newblock {Missing the wealthy in the HFCS: micro problems with macro
  implications}.
\newblock {ECB Working Paper Series} No 2163, European Central Bank.

\bibitem[Chambers and Ren, 2004]{chambers2004outlier}
Chambers, R.~L. and Ren, R. (2004).
\newblock Outlier robust imputation of survey data.
\newblock {\em The Proceedings of the American Statistical Association}, pages
  3336--3344.

\bibitem[Coibion et~al., 2017]{COIBION201770}
Coibion, O., Gorodnichenko, Y., Kueng, L., and Silvia, J. (2017).
\newblock Innocent bystanders? monetary policy and inequality.
\newblock {\em Journal of Monetary Economics}, 88:70 -- 89.

\bibitem[Colciago et~al., 2019]{joes}
Colciago, A., Samarina, A., and de~Haan, J. (2019).
\newblock Central bank policies and income and wealth inequality: A survey.
\newblock {\em Journal of Economic Surveys}, 0(0).

\bibitem[D'Alessio and Neri, 2015]{Neri15}
D'Alessio, G. and Neri, A. (2015).
\newblock {Income and wealth sample estimates consistent with macro aggregates:
  some experiments}.
\newblock Questioni di Economia e Finanza (Occasional Papers) 272, Bank of
  Italy, Economic Research and International Relations Area.

\bibitem[Davison and Smith, 1990]{DavisonSmith1990}
Davison, A.~C. and Smith, R.~L. (1990).
\newblock Models for exceedances over high thresholds.
\newblock {\em Journal of the Royal Statistical Society. Series B
  (Methodological)}, 52(3):393--442.

\bibitem[Deville and S{\"a}rndal, 1992]{DevilleSarndal92}
Deville, J.-C. and S{\"a}rndal, C.-E. (1992).
\newblock Calibration estimators in survey sampling.
\newblock {\em Journal of the American Statistical Association},
  87(418):376--382.

\bibitem[{EG-LMM}, 2017]{eglmm}
{EG-LMM} (2017).
\newblock Understanding, quantifying and explaining the differences between
  macro and micro data of household wealth: Final report.
\newblock mimeo, European Central Bank.

\bibitem[{Eurosystem Household Finance and Consumption Network},
  2009]{hfcs2009}
{Eurosystem Household Finance and Consumption Network} (2009).
\newblock Survey data on household finance and consumption research: summary
  and policy use.
\newblock {ECB Occasional Paper Series} No 100, European Central Bank.

\bibitem[{Eurosystem Household Finance and Consumption Network},
  2020]{hfcs2020a}
{Eurosystem Household Finance and Consumption Network} (2020).
\newblock The household finance and consumption survey:methodological report
  for the 2017 wave.
\newblock {ECB Statistics Paper Series} No 35, European Central Bank.

\bibitem[{Expert Group on Linking macro and micro data}, 2020]{egdna}
{Expert Group on Linking macro and micro data} (2020).
\newblock Understanding household wealth: linking macro and micro data to
  produce distributional financial accounts.
\newblock Statistics Paper Series~37, European Central Bank.

\bibitem[Fesseau and Mattonetti, 2013]{oecd2013}
Fesseau, M. and Mattonetti, M.~L. (2013).
\newblock Distributional measures across household groups in a national
  accounts framework.
\newblock OECD Statistics Working Papers No. 2013/08, Organisation for Economic
  Co-Operation and Development ({OECD}).

\bibitem[Fr{\'{e}}meaux and Leturcq, 2020]{Frmeaux2020}
Fr{\'{e}}meaux, N. and Leturcq, M. (2020).
\newblock Inequalities and the individualization of wealth.
\newblock {\em Journal of Public Economics}, 184:104145.

\bibitem[Gabaix and Ibragimov, 2011]{Gabaix2011}
Gabaix, X. and Ibragimov, R. (2011).
\newblock Rank - 1 / 2: A simple way to improve the ols estimation of tail
  exponents.
\newblock {\em Journal of Business \& Economic Statistics}, 29(1):24--39.

\bibitem[Garbinti et~al., 2018]{gp2018}
Garbinti, B., Goupille-Lebret, J., and Piketty, T. (2018).
\newblock Income inequality in france, 1900{\textendash}2014: Evidence from
  distributional national accounts ({DINA}).
\newblock {\em Journal of Public Economics}, 162:63--77.

\bibitem[Garbinti et~al., 2020]{garbinti2017accounting}
Garbinti, B., Goupille-Lebret, J., and Piketty, T. (2020).
\newblock {Accounting for Wealth-Inequality Dynamics: Methods, Estimates, and
  Simulations for France}.
\newblock {\em Journal of the European Economic Association}.

\bibitem[Guiso et~al., 2005]{GPV}
Guiso, L., Paiella, M., and Visco, I. (2005).
\newblock Do capital gains affect consumption? estimates of wealth effects from
  italian households’ behavior.
\newblock {\em Long-run Growth and Short-run Stabilization: Essays in Memory of
  Albert Ando}.

\bibitem[Haziza et~al., 2017]{haziza2017construction}
Haziza, D., Beaumont, J.-F., et~al. (2017).
\newblock Construction of weights in surveys: A review.
\newblock {\em Statistical Science}, 32(2):206--226.

\bibitem[Kennickell, 2008]{AK2008}
Kennickell, A. (2008).
\newblock The role of over-sampling of the wealthy in the survey of consumer
  finances.
\newblock {\em Irving Fisher Committee Bulletin}, 28.

\bibitem[Kennickell, 2019]{Ken19}
Kennickell, A. (2019).
\newblock The tail that wags: differences in effective right tail coverage and
  estimates of wealth inequality.
\newblock {\em The Journal of Economic Inequality}.

\bibitem[Langousis et~al., 2016]{Langousis2016}
Langousis, A., Mamalakis, A., Puliga, M., and Deidda, R. (2016).
\newblock Threshold detection for the generalized {Pareto} distribution: Review
  of representative methods and application to the {NOAA} {NCDC} daily rainfall
  database.
\newblock {\em Water Resources Research}, 52(4):2659--2681.

\bibitem[Little and Vartivarian, 2005]{little2005does}
Little, R.~J. and Vartivarian, S. (2005).
\newblock Does weighting for nonresponse increase the variance of survey means?
\newblock {\em Survey Methodology}.

\bibitem[Michelangeli and Rampazzi, 2016]{qef_369_16}
Michelangeli, V. and Rampazzi, C. (2016).
\newblock {Indicators of financial vulnerability: a household level study}.
\newblock Questioni di Economia e Finanza (Occasional Papers) 369, Bank of
  Italy, Economic Research and International Relations Area.

\bibitem[Paiella, 2007]{Pa2007}
Paiella, M. (2007).
\newblock {Does wealth affect consumption? Evidence for Italy}.
\newblock {\em Journal of Macroeconomics}, 29(1):189--205.

\bibitem[Ranalli and Neri, 2011]{Ran11}
Ranalli, M.~G. and Neri, A. (2011).
\newblock {To misreport or not to report?, The case of the Italian survey on
  household income and wealth}.
\newblock {\em Statistics in Transition new series}, 12(2):281--300.

\bibitem[Rubin, 1976]{Rubin76}
Rubin, D.~B. (1976).
\newblock Inference and missing data.
\newblock {\em Biometrika}, 63(3):581--592.

\bibitem[Rubin, 1987]{Rubin1987}
Rubin, D.~B., editor (1987).
\newblock {\em Multiple Imputation for Nonresponse in Surveys}.
\newblock John Wiley {\&} Sons, Inc.

\bibitem[Saez and Zucman, 2016]{saez16}
Saez, E. and Zucman, G. (2016).
\newblock Wealth inequality in the united states since 1913: Evidence from
  capitalized income tax data*.
\newblock {\em The Quarterly Journal of Economics}, 131:qjw004.

\bibitem[S{\"a}rndal, 2007]{sarndal2007calibration}
S{\"a}rndal, C.-E. (2007).
\newblock The calibration approach in survey theory and practice.
\newblock {\em Survey Methodology}, page~99.

\bibitem[S\"{a}rndal and Lundstr\"{o}m, 2005]{Sarndal2005}
S\"{a}rndal, C.-E. and Lundstr\"{o}m, S. (2005).
\newblock {\em Estimation in Surveys with Nonresponse}.
\newblock John Wiley {\&} Sons, Ltd.

\bibitem[Schr\"{o}der et~al., 2019]{Schrder2019}
Schr\"{o}der, C., Bartels, C., Grabka, M.~M., K\"{o}nig, J., Kroh, M., and
  Siegers, R. (2019).
\newblock A novel sampling strategy for surveying high net-worth
  individuals{\textemdash}a pretest application using the socio-economic panel.
\newblock {\em Review of Income and Wealth}.

\bibitem[Vermeulen, 2016]{Vermeulen2016}
Vermeulen, P. (2016).
\newblock Estimating the top tail of the wealth distribution.
\newblock {\em American Economic Review}, 106(5):646--50.

\bibitem[Vermeulen, 2018]{vermeulen2018fat}
Vermeulen, P. (2018).
\newblock How fat is the top tail of the wealth distribution?
\newblock {\em Review of Income and Wealth}, 64(2):357--387.

\bibitem[Waltl, 2018]{Waltl2018}
Waltl, S. (2018).
\newblock {Multidimensional Wealth Inequality: A Hybrid Approach toward
  Distributional National Accounts in {E}urope}.
\newblock In {\em Proc.\ 35th IARIW General Conference (IARIW 2018)}.

\bibitem[Yang, 1978]{Yang1978}
Yang, G.~L. (1978).
\newblock Estimation of a biometric function.
\newblock {\em The Annals of Statistics}, 6(1):112--116.

\end{thebibliography}
